\documentclass[reprint,twocolumn,superscriptaddress,showpacs,nofootinbib,notitlepage,preprintnumbers]{revtex4-2}

\usepackage{graphicx}
\usepackage{latexsym,amsmath,amssymb,lmodern,float,url}
\usepackage{natbib}
\usepackage{xcolor}
\usepackage{microtype}

\usepackage[colorlinks=true,backref=false, linktocpage=true, citecolor=blue,urlcolor=blue,linkcolor=blue,pdfpagemode=UseOutlines]{hyperref}

\def\Re {\operatorname{{Re}}}
\def\Im {\operatorname{{Im}}}

\newcommand{\eq}[1]{Eq.~(\ref{#1})}

\begin{document}
\preprint{INT-PUB-22-060}
\title{Deep Learning of Fermion Sign Fluctuations}

\author{Scott Lawrence}
\email{scott.lawrence-1@colorado.edu}
\affiliation{Department of Physics, University of Colorado, Boulder, CO 80309, USA}
\author{Yukari Yamauchi}
\email{yyukari@umd.edu}
\affiliation{Institute for Nuclear Theory, University of Washington, Seattle, WA 98195, USA}
\affiliation{Department of Physics, University of Maryland, College Park, MD 20742, USA}
\date{\today}

\begin{abstract}
We describe a procedure for alleviating the fermion sign problem in which phase fluctuations are explicitly subtracted from the Boltzmann factor. Several ans\"atze for fluctuations are designed and compared. In the absence of a sufficiently high-quality ansatz, a neural network can be trained to parameterize the fluctuations. Demonstrating on the staggered Thirring model in $1+1$ dimensions, we examine the performance of this method as deeper neural networks are used, and in conjunction with the well-studied contour deformation methods.
\end{abstract}

\maketitle

\section{Introduction}

Lattice Monte Carlo methods are the main computational tool used today for strongly coupled field theories. The partition function of the lattice system is expressed as an integral, taken over a space of field configurations $\phi$, of the exponential of the action $e^{-S(\phi)}$. Expectation values---derivatives of the logarithm of the partition function---thus have the form $\langle \mathcal O\rangle = \frac{\int\mathcal{D}\phi\, e^{-S(\phi)} \mathcal O}{\int\mathcal{D}\phi\, e^{-S(\phi)}}$. Where the action is real, field-theoretic expectation values are therefore equal to expectation values over a probability distribution proportional to $e^{-S}$, and can be efficiently computed via importance sampling.

For relativistic fermionic theories at finite chemical potential, the action used for sampling generically has a nonzero imaginary part. As a result, the Boltzmann factor $e^{-S}$ is complex (or, at least, negative), and cannot be treated as a probability distribution. This \emph{fermion sign problem} is the chief obstacle to the use of lattice Monte Carlo methods to study relativistic field theories at non-zero fermion density, including dense QCD matter.

A wide variety of methods have attempted to alleviate, eliminate, or entirely circumvent the exponential scaling of the average phase. Among them are complex Langevin~\cite{Aarts:2008rr}, the density of states method~\cite{Langfeld:2016mct}, canonical methods~\cite{Alexandru:2005ix,deForcrand:2006ec}, reweighting methods~\cite{Fodor:2001au}, series expansions in the chemical potential~\cite{Allton:2002zi}, fermion bags~\cite{Chandrasekharan:2013rpa}, analytic continuation from imaginary chemical potentials~\cite{deForcrand:2006pv}, and finally the contour deformation methods~\cite{Alexandru:2020wrj}.

The most straightforward way of dealing with a sign problem, without substantially modifying the Monte Carlo approach, is termed \emph{reweighting}. A \emph{quenched} expectation value is defined as an expectation value taken over a distribution that ignores the sign fluctuations:
\begin{equation}
\langle \mathcal O \rangle_Q
\equiv
\frac{\int \mathcal D\phi\,|e^{-S(\phi)}| \mathcal O(\phi)}{\int \mathcal D\phi\,|e^{-S(\phi)}|}\text.
\end{equation}
Such quenched expectation values are not of direct physical interest. However, physical expectation values can now be extracted as ratios of quenched expectation values:
\begin{equation}
\langle \mathcal O \rangle
=
\frac
{\langle \mathcal O e^{-i \Im S} \rangle_Q}
{\langle e^{-i \Im S} \rangle_Q}
\text.
\end{equation}
This method is frequently not usable by itself, as the denominator (termed the \emph{average phase}) is generically exponentially small in the spacetime volume of the lattice. As a result, exponentially many samples are required with this method to reliably distinguish any observable from infinity. Nevertheless, the technique of reweighting forms the backbone of many methods for approaching the sign problem, including the subject of this work. We will first alleviate the sign problem to make the average phase manageable, and then treat what remains as above.

This paper begins from the observation that for any model whose Boltzmann factor $e^{-S(\phi)}$ exhibits phase fluctuations, those fluctuations can be entirely removed by subtracting from the Boltzmann factor a function $g(\phi)$ which integrates to zero (which we term a \emph{subtraction}). This modification does not affect the partition function, and therefore leaves suitably defined expectation values invariant as well. The function $g(\phi)$ may be thought of as capturing the phase fluctuations in the Boltzmann factor.

In order for this observation to be useful, we must be able both to find good subtractions $g(\phi)$, and to efficiently guarantee that they do, in fact, integrate to $0$. A first step in this direction was taken in~\cite{Lawrence:2020kyw}, where subtractions were obtained by expanding the Boltzmann factor in some parameter (e.g.~a small coupling). Since the terms in such an expansion could be analytically integrated, it was possible to construct functions guaranteed to integrate to $0$. In effect, this amounted to explicitly removing from the Boltzmann factor all contributions to the sign problem at some fixed order in perturbation theory.

In this paper we provide a more general prescription. Rather than directly search for functions $g(\phi)$, we will construct and optimize a vector field $v_i(\phi)$---that is, a function from field configurations $\phi$ to elements of the tangent space to the space of field configurations at $\phi$. (For example, in the case of single-component scalar fields on a lattice with $N$ sites, the space of field configurations is $\mathbb R^N$ and a vector field is a function $\mathbb R^N \rightarrow \mathbb R^N$.) The subtraction itself is obtained as $g(\phi) \equiv \nabla\cdot v(\phi)$, and is guaranteed to have a vanishing integral by the divergence theorem as long as either the field space is compact, or $v$ decays sufficiently quickly at infinity. The vector field can then be set to any convenient ansatz---in Section~\ref{sec:learning}, a deep neural network---and optimized. At worst, the divergence of the chosen ansatz does not correlate well with the phase fluctuations of the Boltzmann factor, and the average phase is not improved. Regardless of the performance of the chosen ansatz, no approximation has been made, and the result of the final Monte Carlo calculation may be trusted (to within statistical errors).

Early work along these lines is found in~\cite{Doi:2017gmk}, where similar techniques are used to reduce a sign problem in a one-dimensional integral. This line of work was originally motivated by a desire to alleviate certain difficulties encountered by complex Langevin simulations~\cite{Ambjorn:1986fz,Tsutsui:2015tua}.

This approach is closely related to the method of optimized contour deformations (see~\cite{Alexandru:2020wrj} for a review of the whole family of contour deformation methods). There, the original domain of integration in the path integral is viewed as a half-dimension integration contour in a larger complex space. Cauchy's theorem is then invoked to deform this domain of integration without the partition function or any physical observable being modified. This deformation does, however, modify the average phase, and the contour can be optimized via gradient descent to maximize the resulting average phase~\cite{Ohnishi:2019ljc,Kashiwa:2019lkv,Alexandru:2018ddf,Alexandru:2018fqp}. Because the deformed contour is typically parameterized (for computational reasons) by the original contour, such contour deformations can be re-expressed as subtractions of the form above. This correspondence will allow us to apply some of what is known about contour deformations to developing subtractions.

Two key differences separate the approach described here from those based on contour deformations. First, contour deformations that entirely remove the sign problem do not always exist (and in fact, the contour deformation method is entirely inapplicable to theories in which the Boltzmann factor is always real, but fluctuates in sign), while perfect subtractions provably do. Second, contour deformations have yet to be applied to path integrals with discrete variables, while subtraction-based methods are able to address such models naturally. Both these aspects are discussed in more detail in later sections of this work.

In Section~\ref{sec:model} we describe the lattice Thirring model (and its quantum mechanical generalization), which has properties fairly representative of other fermion sign problems and will be used as a testbed throughout this paper. Section~\ref{sec:subtractions} details the method proposed to alleviate the sign problem in detail, and shows how it relates to the longstanding approach of contour deformations. Strategies for analytically constructing subtractions are described and compared in Section~\ref{sec:analytic}, and then deep learning approaches are used in Section~\ref{sec:learning}. Section~\ref{sec:deformations} shows how the method proposed in this work can be combined with contour deformations. Finally, we conclude in Section~\ref{sec:discussion} with a discussion of future avenues of work.

All code used in the simulations discussed in this paper is available online~\cite{code}.

%%%%%%%%%%%%%%%%%%%%%%%%%%%%%%%%%%%%%%%%%%%%%%%%%%%%%
\section{Thirring model}\label{sec:model}

This paper uses a staggered discretization of the Thirring model~\cite{thirring1958soluble} in $1+1$ dimensions as a testbed for the method proposed to alleviate the sign problem. The lattice action of this model is~\cite{Alexandru:2016ejd}
\begin{equation}\label{eq:thirring1+1}
S = \sum_{x,\nu=0,1} \frac{2}{g^2} (1 - \cos A_\nu(x)) - \log \det K[A]
\text,
\end{equation}
The fields $A_\mu(x)$ are valued on $[0,2\pi)$; equivalently, they are the logarithms of $U(1)$-valued link fields. The Dirac matrix is defined by
\begin{widetext}
\begin{equation}
K[A]_{xy} = m \delta_{xy} + \frac 1 2 \sum_{\nu=0,1} \eta_\nu e^{i A_\nu(x) + \mu \delta_{\nu,0}} \delta_{x+\nu,y}
- \eta_\nu e^{-i A_\nu(y) - \mu \delta_{\nu,0}} \delta_{y+\nu,x}
\text.
\end{equation}
The staggered fermions are defined by $\eta_0 = (-1)^{\delta_{0,x_0}}$ and $\eta_1 = (-1)^{x_0}$. Indices $x,y$ refer to sites on a lattice of dimension $\beta \times L$. The bare mass $m$ and coupling $g^2$, in this paper always given in lattice units (i.e.~the lattice spacing is $1$), determine the physical fermion mass and strength of coupling of the theory.

The low-lying spectrum of this lattice Thirring model consists of two flavors of fermion, and a set of bosonic bound states each of a fermion and an antifermion. With bare lattice parameters of $m=0.05$ and $g^2 = 1.0$, the lightest fermion mass is measured to be $a m_F = 0.35(2)$, and the lightest boson mass $a m_B = 0.33(1)$. That the two masses are nearly equal indicates that we are looking at a regime of strong coupling. These bare couplings will be used throughout this paper.

It is occasionally convenient to fall back to a $0+1$-dimensional generalization of the Thirring model, described in~\cite{Alexandru:2015xva,Alexandru:2015sua}.
The action is of essentially the same form, but now with a one-dimensional lattice geometry and one degree of freedom $A(t)$ per site (i.e.~the summation over $\nu$ is removed):
\begin{equation}
S_{\mathrm{mechanics}} = \sum_{t} \frac{1}{2g^2} (1 - \cos A(t)) - \log \det K[A]
\end{equation}
The Dirac matrix $K[A]$ is given by
\begin{equation}\label{mechanics}
K[A]_{t t'} = \frac 1 2
\big[
e^{\mu + i A(t)} \delta_{(t+1)t'}
- e^{-\mu-iA(t')} \delta_{(t'+1)t}
- e^{\mu + i A(t)} \delta_{tN}\delta_{t'1}
+ e^{-\mu-iA(t')} \delta_{t1}\delta_{t'N}
\big] + m \delta_{tt'}
\text.
\end{equation}
\end{widetext}
These formulations of the Thirring model have frequently been used as a testbed for methods to alleviate the sign problem~\cite{Alexandru:2015xva,Alexandru:2015sua,Alexandru:2017czx,Alexandru:2018fqp,Aarts:2008rr,Fujii:2015vha,Pawlowski:2013pje}. Most relevantly here, the Thirring model has played a central role in the development of contour deformation methods, and therefore a good deal is known about how its sign problem can be mitigated via contour deformations.

The earliest contours studied in the context of the $1+1$-dimensional Thirring model were defined by the holomorphic gradient flow. Each point $A$ on the initial, real surface of integration ($\mathbb T^{2\beta L}$, henceforth referred to as the ``real plane'') is evolved according to the first-order differential equation
\begin{equation}
\frac{d A_\mu(x)}{dt} = \left(\frac{\partial S}{\partial A_\mu(x)}\right)^*
\text.
\end{equation}
For a fixed flow time $t$, this equation defines an integration contour $\mathcal M_t \subset (\mathbb C/\mathbb Z)^{2 \beta L}$ in the same homology class as the real plane, which may be hoped to improve the sign problem.

In fact, because the action of the Thirring model has phase fluctuations on the real plane, it can be proved~\cite{Lawrence:2021izu} that there is some sufficiently small flow time $t$ for which the average phase is improved by the flow. Moreover, in practice long flow times were found to be an effective method for alleviating the sign problem~\cite{Alexandru:2016ejd}.

Little is known about the contours $\mathcal M_t$ (for $t \ne 0,\infty$) other than a host of algorithms for performing contour integrals along them. A separate class of contours was investigated in~\cite{Alexandru:2018fqp,Alexandru:2018ddf}, and found to be competitive with the holomorphic gradient flow. These contour deformations are defined by the relations
\begin{eqnarray}
\Im A_0(x) &&= \sum_{n=0}^{\infty} c_n \cos \big(n \Re A_0(x)\big)
\text{, and }\\ \Im A_1(x) &&= 0\nonumber
\text.
\end{eqnarray}
Because of the simple form of this ansatz---in particular, because the determinant of the Jacobian can be computed in linear time---a Monte Carlo integrating along this contour typically outperforms one integrating along $\mathcal M_t$, even where the average phase of the latter might be larger.

A \emph{post-hoc} rationalization for this ansatz was put forward in~\cite{Lawrence:2018mve}: in the limit of large chemical potential $a\mu \gg 1$, the partition function factorizes and this form of contour deformation exactly solves the sign problem. At any finite chemical potential, of course, the sign problem remains and is exponential in the volume.

Some of the algorithms discussed in the following sections suffer due to the presence of zeros of the Boltzmann factor. This can be alleviated by working with an integration contour which avoids those zeros. One such contour is easy to construct, defined by a simple shift of all time-like vectors
\begin{equation}\label{eq:shift}
\tilde A_0 = A_0 + i \mu
\end{equation}
and $\tilde A_1 = A_1$ at each lattice site. This contour has the effect of placing the chemical potential into the ``bosonic'' part of the action (resulting in a term of the form $\cos A_0 - i\mu$), while removing it entirely from the Dirac matrix. Because the $\mu = 0$ Dirac matrix has no zeros for real values of the fields, this results in a Boltzmann factor and action which are always finite.

%%%%%%%%%%%%%%%%%%%%%%%%%%%%%%%%%%%%%%%%%%%%%%%%%%%%%
\section{Subtractions}\label{sec:subtractions}
Consider a general model with action $S$---a function of some fields $\phi$ which will be elided for brevity. The partition function of this system is $Z = \int e^{-S}$, and we can define a quenched partition function by $\int |e^{-S}|$, which is different whenever there is a sign problem.

Let $g$ be any function of the fields which integrates to zero. Constructing a modified Boltzmann factor $e^{-S_{\mathrm{sub}}} \equiv e^{-S} - g$, note that the partition function is not changed: $Z = \int e^{-S_{\mathrm{sub}}}$. Physical expectation values, obtained by differentiating the partition function with respect to appropriate source fields, consequently do not change either. However, the quenched partition function (and therefore the average phase) is generically modified by this transformation:
\begin{equation}
Z_{Q,\mathrm{sub}} = 
\int |e^{-S_{\mathrm{sub}}}|
=
\int |e^{-S} - g|
\ne \int |e^{-S}|
\text.
\end{equation}
We will refer to the function $g$ as a ``subtraction''. The observation that the quenched partition function can be modified while leaving physical expectation values untouched suggests that a well-constructed $g$ might provide a way around a fermion sign problem.

As noted in~\cite{Lawrence:2020kyw}, for any Boltzmann factor, a function $g$ can always be found that entirely removes the sign problem. The most straightforward construction, in the case of a compact field space, is:
\begin{equation}\label{exact}
g_{\mathrm{exact}}(\phi) = e^{-S}(\phi) - \frac{\int \mathcal D \phi' \, e^{-S}(\phi')}{\int \mathcal D \phi'}
\text.
\end{equation}
This construction is far from unique. For example, let $h$ be any function supported on a region of field space where the magnitude of the Boltzmann factor is bounded from below by some non-zero constant. Then there exists some sufficiently small $\epsilon$ for which $\tilde g = g_{\mathrm{exact}} + \epsilon h$ is also an exact subtraction, in the sense that $e^{-S} - g_{\mathrm{exact}} - \epsilon h$ has no phase fluctuations.

Although perfect subtractions are not unique objects, they form a very small subset of the space of all functions that integrate to zero. Recall that the average phase $\langle \sigma \rangle \equiv \frac{Z}{Z_Q}$ is exponentially small in spacetime volume: $\langle \sigma \rangle \sim e^{-V}$. As a result, the typical magnitude of a subtracted Boltzmann factor must be exponentially smaller than the typical magnitude of the original Boltzmann factor. The subtraction itself must therefore be equal to the original Boltzmann factor within exponential (in $V$) precision.

In order to make use of a subtraction to construct an unbiased Monte Carlo algorithm, we must be able to prove that it integrates to $0$. This may be accomplished by defining $g \equiv \nabla \cdot v$ for some vector field $v$. The resulting subtraction is guaranteed to integrate to $0$ by the divergence theorem. The effective ``subtracted action'' is now defined by
\begin{equation}\label{unscaled}
e^{-\tilde S_v} \equiv e^{-S} - \nabla \cdot v\text.
\end{equation}

The vector field $v$ used above will typically be required to be of order $e^{-S}$. Since the fluctuations of the action are generally algebraic in the spacetime lattice volume, the fluctuations in the magnitude of the vector field will be exponential in the size of the system. Particularly when we turn to using machine learning methods to train the vector field in Section~\ref{sec:learning}, this will be difficult to accomplish. Therefore, we find it useful to define a ``scaled'' vector field by $v = e^{-S} u$. The vector field $u$ may now be generically of order $1$. The effective subtracted action is now given by
\begin{equation}\label{scaled}
e^{-S_u} \equiv e^{-S} - \nabla \cdot e^{-S} u
=
e^{-S} \left(1
+ u \cdot \nabla S
-\nabla \cdot u
\right)
\text.
\end{equation}
A large part of the appeal of modifying a Boltzmann factor with subtractions---rather than, for instance, contour deformations---is that following the discussion above, it is easy to prove the existence of a ``perfect'' subtraction entirely removing the sign problem. Unfortunately, this scaling trick, while numerically valuable, can break this property. When the Boltzmann factor $e^{-S}$ has a zero, any finite vector field $u$ will result in a scaled vector field $v$ with the same structure of zeros. This is a nontrivial constraint that means that some potential subtractions cannot be represented with this method.

This is a technical obstacle with various possible workarounds. In this paper, it is sufficient to shift the integration contour as per the discussion around Eq.~(\ref{eq:shift}) to obtain a Boltzmann factor that has no zeros. For other models, other tricks may be required (e.g.~to allow the vector field $u$ to have well-controlled divergences), and in still other cases, the issue does not appear (scalar field theories have no zeros of the Boltzmann factor).

For a subtraction to be of practical use, we must also specify a way to compute an observable via Monte Carlo. The expectation value of a function $\mathcal O(\phi)$ will be different over the original distribution $e^{-S(\phi)}$ and the subtracted distribution $e^{-\tilde S_v(\phi)}$, as a consequence of the fact that although $\int \nabla\cdot v$ vanishes, there is no guarantee that $0 = \int \mathcal O \nabla \cdot v$. One prominent exception deserves to be mentioned before we move on: if $v$ is trained with the constraint that it be orthogonal to $\nabla \mathcal O$ for some observable $\mathcal O$, then the function $\mathcal O(\phi)$ will have the same expectation value before and after the subtraction. We do not pursue this approach further here\footnote{This constraint has several obvious drawbacks, in particular that it does not scale well to the computation of multiple observables, and that there is no longer a guarantee that a perfect subtraction is available. One less obvious (but serious in practice) drawback is that, for fermionic theories, additional gradients of the fermion determinant must be computed.}.

An expression that always gives the correct expectation value for an observable $\mathcal O$, without requiring any constraints on the vector field used, is
\begin{equation}\label{obs-bad}
\langle \mathcal O \rangle
=
\frac{\int e^{-\tilde S_v} \frac{e^{-S} \mathcal O}{e^{-\tilde S_v}}}{\int e^{-\tilde S_v}}
=
\Big\langle \frac{e^{-S}\mathcal O}{e^{-\tilde S_v}}\Big\rangle_{\tilde S_v}
\text.
\end{equation}
While appealingly simple, the corresponding algorithm is also not useful in practice, as the expectation value with respect to the subtracted distribution typically has a signal-to-noise problem at least as severe as original sign problem. This can be seen most clearly by considering the case of the trivial observable $\mathcal O = 1$, and a perfect subtraction defined by \eq{exact}. In that case, the sampling is being performed with respect to a flat distribution over field space, and the average phase is being computed by a Monte Carlo without importance sampling.

Instead, working from a subtraction of the form of \eq{scaled}, from the numerator) the expectation value of $\mathcal O$ is obtained according to
\begin{equation}\label{obs-good}
\langle \mathcal O\rangle
= \Big\langle \mathcal O - \frac{e^{-S} u \cdot \nabla \mathcal O}{e^{-S_u}}\Big\rangle_{S_u}\text.
\end{equation}
Empirically, we find that this form of the numerator does not have a severe sign-to-noise problem. However, we do not have a good analytic argument for why this should be the case, and it remains unclear if this expression is in any sense optimal.

The fact that there is no unique expression for the expectation value follows from the observation that, just as we modified the denominator by subtracting the divergence of an arbitrary vector field, the numerator can be modified in the same way. Different subtractions in the \emph{numerator} yield different expressions for the same expectation value, all taken over the same subtracted distribution. The expression of Eq.~(\ref{obs-good}) may be obtained by subtracting $\nabla\cdot (v \mathcal O)$ from the numerator. This perspective suggests that, in future work, the numerator subtractions might be optimized separately from the one in the denominator.

To conclude this section, it is instructive to consider the connections between contour deformation methods and the subtraction-based methods described here. First, as mentioned in the introduction, every contour deformation parameterized by the real plane\footnote{This includes all contour deformations considered in the literature on alleviating sign problems, but not all possible contour deformations: there is no need for an integration contour to be homeomorphic to the real plane in order to be in the same homology class.} necessarily results in a subtraction. In the case of a field $\phi \in \mathbb R^N$ complexified to $\tilde\phi(\phi) \in \mathbb C^N$, this subtraction is defined by:
\begin{equation}
g_{\mathrm{sub}}(\phi)
= e^{-S(\tilde\phi(\phi))} \det \frac{\partial \tilde\phi}{\partial \phi} - e^{-S(\phi)}
\end{equation}
That $g_{\mathrm{sub}}$ integrates to $0$ may be established from Cauchy's integral theorem.

The converse does not hold: not all subtractions can be formulated as contour deformations. A simple example is given by a one-site lattice with Boltzmann factor $e^{-S(\theta)} = \cos\theta + \epsilon$. The sign problem is entirely removed by subtracting off $\cos\theta$, but there is no contour deformation that increases the average phase~\cite{Lawrence:2020irw}.

When using a subtraction directly, one must make a choice of how expectation values are to be computed (e.g.~\eq{obs-bad} \emph{versus} \eq{obs-good}). The expectation value obtained does not depend on this choice, but the efficiency of the algorithm does. When performing a contour deformation, no such choice is available. Rather, the new form of the observable is determined by analytic continuation of the original observable to the deformed contour.

In the limit of small contour deformations, the behavior of both the observable and the Boltzmann factor is determined from their respective first derivatives alone. In this limit, the contour deformation algorithm precisely matches the prescription of \eq{obs-good}, as we now demonstrate. Consider an integration contour, parameterized by the real plane, defined by a function $\tilde \phi:\mathbb R^N \rightarrow \mathbb C^N$:
\begin{equation}\label{contour-infinitesimal}
\tilde\phi_i(\phi) = \phi_i - u_i(\phi)
\end{equation}
We are interested in the limit of small $u$, and will expand all expressions to first order. Note that at this order, $u$ may be taken to be purely imaginary without loss of generality.

The effective Boltzmann factor, expanded to first order in $u$, is
\begin{eqnarray}
e^{-S_{\mathrm{eff}}(\phi)}
&=& e^{-S(\tilde\phi(\phi))} \det \frac{\partial \tilde\phi}{\partial\phi} 
\nonumber\\
&=& e^{-S(\phi)} \left[
1 + u \cdot \nabla S - \nabla \cdot u
\right]
\text.
\end{eqnarray}
The same expansion yields, for the numerator $\mathcal O e^{-S_{\mathrm{eff}}}$:
\begin{equation}
\mathcal O(\tilde\phi(\phi)) e^{-S_{\mathrm{eff}}(\phi)}
= e^{-S_{\mathrm{eff}}(\phi)}\mathcal O(\phi) - e^{-S(\phi)} u \cdot \nabla \mathcal O
\text.
\end{equation}
An infinitesimal contour deformation of the form \eq{contour-infinitesimal} therefore corresponds precisely to an infinitesimal subtraction---with observables evaluated according to \eq{obs-good}---using precisely the same vector field $u$. The two methods differ only at higher orders in the size of the vector field.

To conclude, we sketch the algorithm being proposed (excluding the choice of vector field) in its entirety. Given a suitable vector field defined on field space---$u_i(\phi)$---, a scaled vector $v = e^{-S} u$ is constructed. Noting that the divergence of $v$ vanishes, we can define a subtracted Boltzmann factor $e^{-S_u}$ according to \eq{scaled}. The partition function is unchanged by this transformation. Sampled field configurations are collected via importance sampling according to the new Boltzmann factor. With a set of samples obtained, the expectation value of an operator $\mathcal O(\phi)$ is computed via \eq{obs-good}.

%%%%%%%%%%%%%%%%%%%%%%%%%%%%%%%%%%%%%%%%%%%%%%%%%%%%%
\section{Analytic subtractions}\label{sec:analytic}

In many cases, useful subtractions can be obtained analytically, without needing any machine learning methods. In this section we construct a few candidates and compare their performance. In subsequent sections we will also see that the constructions of this section provide useful information about what sorts of ``feature engineering'' may be helpful for a neural network (i.e.~what functions of the field configuration might be precomputed and passed directly as inputs to the network).

The $0+1$-dimensional generalization of the Thirring model---see \eq{mechanics}---is exactly solvable~\cite{Pawlowski:2014ada}. The partition function is given by
\begin{eqnarray}
Z &=& e^{-\frac{\beta}{2g^2}} 2^{1-\beta}
\Big[
I^\beta_1\Big(\frac{1}{2g^2}\Big) I(\cosh \beta \mu)\nonumber\\
&&+
I^\beta_0\Big(\frac{1}{2g^2}\Big) I(\cosh (\beta \sinh^{-1} m))
\Big]
\text.
\end{eqnarray}
Above, $I_n(\cdot)$ is the modified Bessel function of the first kind. As a result, an exact subtraction is immediately obtained from \eq{exact}.

\begin{figure}
\centering
\includegraphics[width=0.9\linewidth]{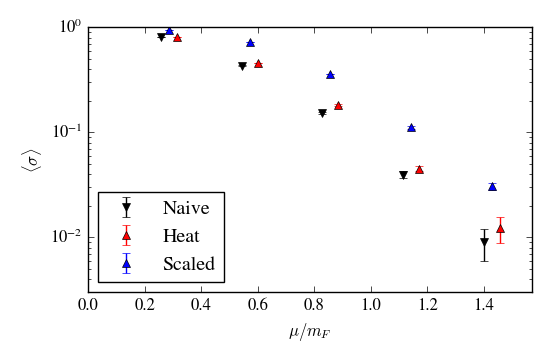}
\caption{Comparison of the performance of analytically constructed subtractions, on a $6 \times 6$ lattice with bare parameters as described in the text. Each data point corresponds to $5 \times 10^{4}$ samples.\label{fig:analytic}}
\end{figure}

For models that aren't exactly solvable (including that of primary interest here, the Thirring model in $1+1$ dimensions), it was suggested in~\cite{Lawrence:2020kyw} to derive a subtraction via an analytically tractable expansion. The heavy-dense expansion, in which $a \mu \gg 1$, is particularly useful for this purpose. Expanding the determinant of the Dirac matrix via the polymer representation~\cite{montvay1997quantum}, the dominant term in this limit is
\begin{equation}
\det K = e^{\beta L\mu} \Big(2^{-\beta L} e^{i \sum_x A_0(x)} + O(e^{-\beta \mu})\Big)
\text.
\end{equation}
This term does not couple different $A$ fields, and the bosonic part of the action does not either. As a result, we can perform the integration of this term analytically, and exactly remove its contribution to the partition function---and therefore the average phase. Because this term is the dominant term at large $\mu$, it represents a substantial contribution to the sign problem, and removing it is a significant improvement. The final heavy-dense subtraction is
\begin{widetext}
\begin{equation}
g_{\mathrm{hd}}(A) = e^{\frac{2}{g^2} \sum_{x,\nu}\cos A_\nu(x)}
2^{-\beta L} e^{\beta L \mu + i \sum_x A_0(x)}
- e^{\beta L \mu} (4 \pi)^{-\beta L} \left(I_0(2/ g^2) I_1(2/ {g^2})\right)^{\beta L}
\text.
\end{equation}
\end{widetext}

We can also obtain a useful subtraction by beginning with the construction of \eq{scaled} and expanding in the limit of small $u$. (In light of the gradient descent-based methods to be discussed in the following section, this technique can be viewed as performing gradient descent in the space of all possible vector fields $u$.) Expanding the subtracted quenched partition function to first order in $u$ yields
\begin{equation}\label{eq:Zqu}
Z_{Q,u} = \int |e^{-S}| \left[
1 + \Re \left(u \cdot \nabla S -\nabla \cdot u\right) + O(u^2)
\right]
\text,
\end{equation}
and as a result the first functional derivative with respect to $u$ is (see Appendix~\ref{app:deriv}) for details)
\begin{equation}\label{sub-u}
\frac{\delta}{\delta u(\phi)}Z_{Q,u}
=
- i \nabla \Im S
\text.
\end{equation}
Performing a single step of gradient descent in the space of possible vector fields $u$ then results in $u \propto \frac{\delta}{\delta u} Z_{Q,u}$, with the constant of proportionality to be determined by numerical optimization. The gradient descent-based algorithm used in practice for this optimization is detailed in the next section, where it is applied to much larger optimization problems with subtractions parameterized by neural networks.

This construction of a subtraction has a particularly close relationship with the ``sign-optimized contour'' approach~\cite{Ohnishi:2019ljc,Kashiwa:2019lkv,Alexandru:2018fqp,Alexandru:2018ddf}. As noted in the previous section, to leading order in the size of the subtraction and contour deformation, these two methods yield identical effective actions. In either case, the effective Boltzmann factor post-subtraction is
\begin{equation}
e^{-S_{\mathrm{eff}}}
=
e^{-S}
+ i \epsilon e^{-S} \left[\nabla^2 \Im S - \nabla S \cdot \nabla \Im S\right]
\text,
\end{equation}
with $\epsilon$ an arbitrary constant to be determined by optimization.

This connection, although conceptually convenient, is cause for some concern. Notably, contour deformations cannot cure certain sign problems, such as that associated with the Boltzmann factor $(\cos \theta + \epsilon)$, or the mean-field Thirring model~\cite{Lawrence:2021izu}. This implies that there are some actions for which a single step of gradient descent parallel to \eq{sub-u} will provide no improvement. In fact, $(\cos \theta + \epsilon)$ is precisely such an example. The imaginary part of the action is locally constant at every non-singular point, and therefore the functional derivative is almost everywhere zero.

In hopes of having a remedy available, we will now consider two other possible subtractions derived from gradient descent. The underlying intuition is that the path taken by gradient descent is dependent on the local metric chosen to define distances on the space being explored. Evaluating a functional derivative with respect to $u$ is a choice---convenient for several reasons, but ultimately arbitrary. Making different choices will yield different ``first-order subtractions'', which may have better performance characteristics, depending on the circumstances.

The first, and perhaps most obvious, alternative is to begin with \eq{unscaled}, and perform gradient descent on the unscaled field $v$. Again expanding the quenched partition function to leading order in $v$, we find that
\begin{equation}\label{eq:Zqv}
Z_{Q,v} = \int |e^{-S}| \left[1 - \Re e^{S} \nabla \cdot v + O(v^2)\right]
\text.
\end{equation}
In this case, taking the functional derivative (see again Appendix~\ref{app:deriv} for details), we obtain a gradient of
\begin{equation}\label{sub-v}
\frac{\delta}{\delta v(\phi)}Z_{Q,v}
= i e^{-i \Im S} \nabla \Im S
\text.
\end{equation}
Once again, this subtraction is not able to make any progress on the sign problem from an action $S = -\log(\cos\theta+\epsilon)$. Fundamentally, this is because we are differentiating only with respect to local impulses of the vector field.

We can do better by evolving the Boltzmann factor according to the heat equation. Define $S_t(\phi)$ to be the solution to the partial differential equation
\begin{equation}
\frac{d}{dt} e^{-S_t(\phi)}
=
- \sum_i \frac{\partial^2}{\partial \phi_i^2} e^{-S_t(\phi)}
\end{equation}
with initial condition $S_0(\phi) = S(\phi)$. In the limit of long times, this must asymptote to a constant action, which necessarily has no sign problem. Meanwhile, because the time-derivative of the Boltzmann factor is a total derivative (with respect to the fields), the partition function itself is never modified by this transformation.

Of course, it is not practical to perform this PDE evolution, which would require exponential resources in the physical volume. Nevertheless, it motivates a subtraction obtained by considering a single time-step of this evolution.
\begin{equation}\label{eq:sub-heat}
e^{-S} \rightarrow e^{-S} - (\Delta t)
\left(\sum_i \frac{\partial^2}{\partial \phi_i^2} e^{-S(\phi)}\right)
\end{equation}
Note that in the case of the action $S = -\log(\cos \theta + \epsilon)$, a single step of gradient descent (with sufficiently large step size) is enough to entirely remove the sign problem.

Figure~\ref{fig:analytic} compares the performance of two of the various subtractions constructed here to the naive sign problem. The subtraction defined from Eq.~(\ref{sub-u}) is referred to as ``scaled'', and that from Eq.~(\ref{eq:sub-heat}) is marked by the label ``heat''. In both cases there is a single free parameter, the magnitude of the subtraction, which has been optimized by stochastic gradient descent. From the figure it is clear that, although both subtractions result in a measurable improvement to the average phase, that of Eq.~(\ref{sub-u}) is considerably superior.

%%%%%%%%%%%%%%%%%%%%%%%%%%%%%%%%%%%%%%%%%%%%%%%%%%%%%
\section{Machine learning}\label{sec:learning}

\begin{figure*}
\centering
\includegraphics[width=0.45\linewidth]{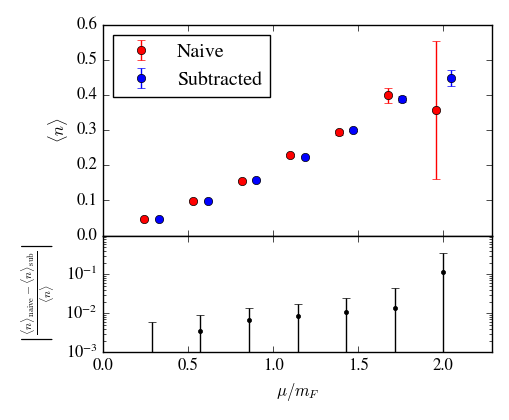}
\hfil
\includegraphics[width=0.45\linewidth]{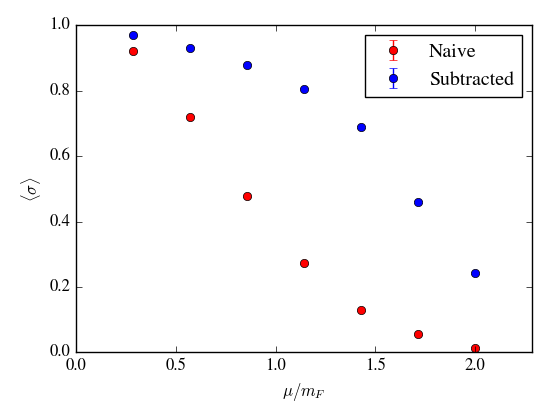}
\caption{Precision test of the correctness of the Monte Carlo on a subtracted Boltzmann factor, comparing number densities computed on a $4\times 4$ lattice at various chemical potentials. Each data point corresponds to $10^5$ samples.\label{fig:correctness}}
\end{figure*}
In this section we apply deep learning to the construction and optimization of a vector field, which will define a subtraction via Eq.~(\ref{scaled}).
A neural network represents the vector field $u$, i.e., its inputs are the field values $A$ (or chosen functions of $A$), and the output is the $V$-dimensional complex vector $u$. Here $V$ is the real dimension of the field space, which is equal to double the number of lattice sites for the lattice Thirring model defined via Eq.~(\ref{eq:thirring1+1}). The divergence $\nabla\cdot (e^{-S} u)$ gives the subtraction function. To optimize the vector field, we apply the same method introduced in~\cite{Alexandru:2018fqp} for training contour deformations. The cost function---strictly speaking, a functional from the space of vector fields $v$ to the non-negative real numbers---is
\begin{equation}\label{eq:cost}
C[u] = -\log \langle \sigma\rangle_u\text.
\end{equation}
As mentioned in Section~\ref{sec:subtractions}, a perfect subtraction always exists, and moreover is far from unique. As any function with an integral of zero may be written as the divergence of some vector field (again, typically in a non-unique way), the cost function attains a global minimum of $C[u] = 0$ at at least one point. In fact, when viewed on the space of all vector fields (rather than merely those representable by a neural network of fixed size), the only local minima and saddle points of the cost function must also be global minima.

Unfortunately (and as discussed briefly in Section~\ref{sec:subtractions}), for numerical reasons it is necessary to work in terms of a scaled vector field $u$; in other words, the subtraction is constructed according to Eq.~(\ref{scaled}) rather than Eq.~(\ref{unscaled}). The learning process performs much better in practice; however, we lose the nice property that we are guaranteed that a perfect subtraction may be obtained. This stems from the fact that the neural network representing $u$ is unable to represent a function with a singularity. For concreteness, consider once again the example of a theory with one degree of freedom, $\theta$, governed by the action $S = - \log (\cos\theta + \epsilon)$. The Boltzmann factor $\cos \theta + \epsilon$ has two zeros, separating the region of positive weight from the region of negative weight. In order for a subtraction $\nabla \cdot v$ to remove the sign problem, it is necessary that positive weight be transported from one region to the other, through those zeros. As a result, we must have $v \ne 0$ where $0 = \cos \theta + \epsilon$. However, if we work in terms of a scaled vector field $u$, so that the subtraction is given (following Eq.~(\ref{scaled})) by $\nabla \cdot (\cos \theta + \epsilon) u$, then this is clearly not possible for $u$ constrained to be finite.

It is unclear how large of a handicap this is in practice. It is likewise unclear if a method exists to restore the guarantee of perfect subtractions, while preserving the nice algorithmic properties of the scaled vector fields. For the remainder of this work we focus on numerical optimization of scaled vector fields, leaving such questions to future work.

The cost function of Eq.~(\ref{eq:cost}) itself is as expensive to compute as any other observable. However, noting that the partition function does not depend on the choice of vector field $u$, the gradient of the cost function with respect to some parameter $\lambda$ defining the vector field is
\begin{equation}\label{eq:gradsigma}
\frac{\partial}{\partial \lambda}C[u(\lambda)]
=
\frac{\partial}{\partial \lambda}\log Z^{(\lambda)}_Q
=
-\langle \Re S \rangle_Q
\text.
\end{equation}
Thus the gradient of the cost function, which is all that is needed in order to apply an optimization algorithm like \textsc{Adam}~\cite{kingma2014adam}, has the form of a quenched expectation value and can therefore be computed relatively efficiently.

The real and imaginary parts of the vector field $u$ are obtained from a multi-layer perceptron---that is, a densely connected neural network. Each hidden layer is given a width of $4V$, where $V$ is the number of lattice sites. The inputs to the network are the sine and cosine of each link variable $A_\mu(x)$---$4V$ values in total. Hidden layers are acted on by the CELU activation function~\cite{barron2017continuously}. Finally, the network has $4V$ outputs, interpreted as the real and imaginary parts of each of $2V$ complex components of a vector.

The training is performed with \textsc{Adam}~\cite{kingma2014adam} according to a scheduled learning rate defined by
\begin{equation}\label{eq:care}
\eta(n) = \begin{cases}
      10^{-4} + \frac{10^{-2}-10^{-4}}{200}n & 1\le n \le 200\\
      10^{-2}\times 0.1^{\frac{n-200}{C\times10^2}} & 200 \le n
    \end{cases} 
\text.
\end{equation}
In other words, for the first $200$ steps of training, the learning rate is increased linearly from $10^{-4}$ to $10^{-2}$; afterwards the learning rate falls exponentially. The training continues until the learning rate drops below $2 \times 10^{-5}$, at which point the process is considered complete. The influence of the free parameter $C$, governing the rate of the exponential decay, is considered in detail later in this section.

At each training step, 100 samples are used to estimate the gradient. As discussed in~\cite{Giordano:2022miv}, it is advantageous to perform multiple steps of gradient descent while re-using the same set of samples. To decide when the samples are stale and a new Monte Carlo should be performed, we define an ``average reweighting'' according to
\begin{equation}\label{eq:R}
R = \frac 1 {100} \sum_k \left|\log \frac{e^{-S(\phi_k)} - \tilde g(\phi_k)}{e^{-S(\phi_k)} - g(\phi_k)}\right|
\text.
\end{equation}
Here $g(\cdot)$ is the subtraction used to perform the Monte Carlo, $\tilde g(\cdot)$ is the current subtraction (after some number of steps of gradient descent), and $\phi_k$ for $k = 1 \ldots 100$ denote the $100$ samples collected. The same set of samples are re-used until either $100$ steps of gradient descent have been taken, or the average reweighting $R$ rises above $0.3$.

To ensure that the path integral modified by the machine-learned subtraction gives the correct physics, we compare the expectation value of the density $n = \frac{1}{\beta V}\log\frac{\partial Z}{\partial\mu}$ with and without the subtraction. As shown in Figure~{\ref{fig:correctness}}, the density on the $4\times 4$ lattice with $m=0.05$, $g^2=1.0$, and varying chemical potential $\mu$ agree at high precision. The subtraction is defined from a neural network as described above, with $2$ internal layers and trained with $C=1.0$. In this demonstration, the average phase and the density are computed with $10^5$ samples from MCMC.

\begin{figure}
\centering
\includegraphics[width=0.9\linewidth]{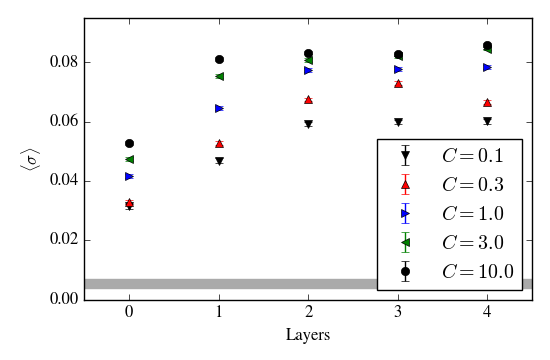}
\caption{Average phase for different network depths and training times. The different data series correspond to increasing the parameter $C$ described in the text; the $x$-axis shows the number of hidden layers. Training time is linear both in the number of layers and $C$. Each data point corresponds to $10^6$ samples; the shaded region shows the $1$-sigma error bars on the average phase with no subtraction applied, computed with $2 \times 10^5$ samples.\label{fig:care}}
\end{figure}

We now consider the dependence of the average phase attained on the size of the network (parameterized by the number of layers) and how it is trained (the parameter $C$ in Eq.~(\ref{eq:care})). We use the same multi-layer perceptron with the CELU activation function as we vary the number of internal layers. The choice of CELU activation function is made so that we avoid the vanishing gradient problem while ensuring that the output vector field be smooth. The training procedure is the same as the previous demonstration in Figure~\ref{fig:correctness} --- 100 samples are used to estimate the gradient at each training step and those samples are reused according to Eq.~(\ref{eq:R}). The training is performed with \textsc{Adam}~\cite{kingma2014adam} according to the scheduled learning rate in Eq.~(\ref{eq:care}). We did use not further techniques for training deep neural networks such as skip connections or dropout. The results are shown in Figure~\ref{fig:care} for a $6\times 6$ lattice with $m=0.05$, $g^2=1.0$, and $\mu=0.5$.

All training procedures attempted showed a substantial improvement in average phase over the naive (subtraction-free) calculation; however, the size of the improvement varied by a factor of $\sim 3$ between a shallow, quickly trained network, and a deep, carefully trained one. That deeper networks systematically outperform shallower ones indicates that the large number of parameters in those ans\"atze is important. However, for the deeper networks (of two or more layers), increasing the training time is typically a more effective way to improve the average phase than increasing the depth, suggesting that future work focus on applying methods to improve the speed of the training rather than the size of the ansatz.

Finally, note the scale on Figure~\ref{fig:care}---the largest network, with the longest training, remained more than a factor of $10$ away from being a perfect subtraction with $\langle\sigma\rangle = 1$. Furthermore, deeper networks with larger training times all cluster around the same value of $\langle\sigma\rangle$, indicating that the training has in some sense converged. This is in tension with the observation made in previous sections, that perfect subtractions do exist. It is unclear why the machine learning procedure shows a clear indication of having reached a barrier at $\langle\sigma\rangle \sim 0.09$.

%%%%%%%%%%%%%%%%%%%%%%%%%%%%%%%%%%%%%%%%%%%%%%%%%%%%%
\section{Contour deformations}\label{sec:deformations}

The intellectual precursor to the method being proposed in this paper was that of contour deformations. The purpose of this section is to show how this method can be combined with the subtractions constructed in this paper to obtain an algorithm yielding superior average phases to either individual method.

As already mentioned, the contour deformation approach suffers from one difficulty above all others: it is unclear whether, and under what conditions, contour deformations exist that can remove the exponential decay of the average phase. In a small number of contexts it is known that contour deformations are not helpful at all. For example, infinitesimal contour deformations are proven not to improve the average phase when the Boltzmann factor on the real plane is real~\cite{Lawrence:2021izu}.

Notwithstanding the uncertainties regarding this approach's general applicability, a small family of contour deformations was proposed for the Thirring model in~\cite{Alexandru:2018fqp} and found to perform sufficiently well to enable calculations of the equation of state on moderately sized $2+1$-dimensional lattices~\cite{Alexandru:2018ddf}. Simple contour deformations have similarly been found effective for lattice scalar $\phi^4$ theory with a complex coupling~\cite{Lawrence:2022afv} or a chemical potential~\cite{Bursa:2018ykf}.

\begin{figure}
\centering
\includegraphics[width=0.9\linewidth]{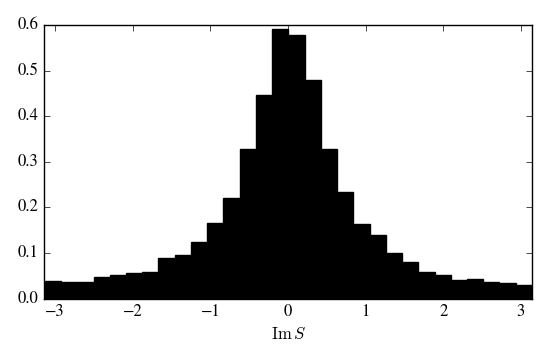}
\caption{Distribution of the imaginary part of the action on the deformed contour of \eq{smile}. The contour is optimized on a $6 \times 6$ lattice at $\mu = 0.5$.\label{fig:smile-histogram}}
\end{figure}

The contour proposed for the Thirring model in~\cite{Alexandru:2018fqp}, defined by the relations
\begin{eqnarray}\label{smile}
\Im A_0(x) &&= \sum_{n=0}^{\infty} c_n \cos \big(n \Re A_0(x)\big)
\text{ and }\\ \Im A_1(x) &&= 0\nonumber
\text,
\end{eqnarray}
is not strictly optimal. To establish this, it is sufficient to consider the distribution of the imaginary part of the action when sampling is performed on the contour with respect to the quenched distribution. As shown in~\cite{Lawrence:2021izu}, on a locally optimal integration contour---defined as one from which no infinitesimal deformation can improve the average phase---the derivative of the imaginary part of the action vanishes everywhere except at zeros of the Boltmann factor. Equivalently, the distribution of $\Im S$ is discrete. A histogram of the imaginary part of the action on a numerically optimized contour of the form of \eq{smile} is shown in Figure~\ref{fig:smile-histogram}. No evidence of multimodality is present, establishing that there are small deformations available along which the average phase would improve.

The existence of small deformations also means that there are infinitesimal subtractions that would improve the average phase. To find one such subtraction, it is sufficient to recall that the ansatz of Eq.~(\ref{scaled}) has a close correspondence with infinitesimal contour deformations: to leading order in the size of the contour deformation, it is equivalent to a subtraction of the form given in that equation.

The use of the ansatz Eq.~(\ref{scaled}) can be naturally combined with the expressive power of neural network via \emph{feature engineering}. Relevant functions, inspired by the form of the ansatz, are pre-computed and fed as inputs to the neural network. For demonstration purposes, we append (to the usual $4V$ inputs described above) the $2V$ gradients of the model's action with respect to the field values $A_\mu(x)$. Each gradient is a complex number, fed into the network as a separate real and imaginary part---therefore there are $4V$ additional inputs. The remaining architecture of the network is unchanged.

\begin{figure}
\centering
\includegraphics[width=0.9\linewidth]{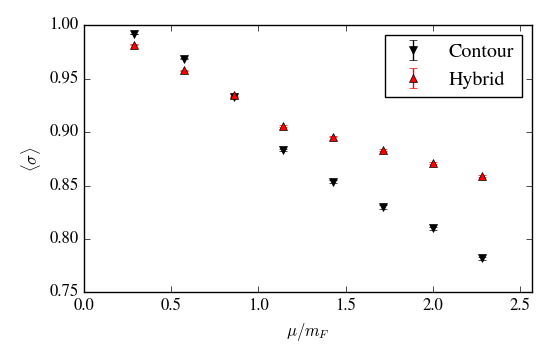}
%\includegraphics[width=0.45\linewidth]{figure6-left.png}
%\hfil
%\includegraphics[width=0.45\linewidth]{figure6-right.png}
\caption{Improvement of the contour deformation of Eq.~(\ref{smile}) via a subtraction. The comparison is performed on a $4 \times 4$ lattice, and each data point corresponds to $10^5$ samples. \label{fig:deformations}}
\end{figure}

Figure~\ref{fig:deformations} compares the average phase on optimized contours of the form of Eq.~(\ref{smile}) with the average phase obtained by using that contour and then applying a subtraction. The subtraction is defined by a zero-layer neural network, with engineered features as described above, and trained with $C=3.0$. For small values of $\mu$, the ``hybrid'' approach yields a \emph{lower} average phase than the contour deformation alone. This is attributable to noise in the stochastic gradient descent, and this feature would be removed with a larger value of $C$.

Note that, since the features input to the neural network are the first derivatives of the action, computing the subtraction involves computing the matrix of \emph{second} derivatives, due to the additional divergence taken in constructing the subtraction. We find that this is a serious drawback of the feature engineering approach in practice. Natural choices of features all have reference to the fermion determinant, and therefore higher derivatives of that object are required.

%%%%%%%%%%%%%%%%%%%%%%%%%%%%%%%%%%%%%%%%%%%%%%%%%%%%%
\section{Discussion}\label{sec:discussion}

We have demonstrated a family of methods for alleviating the fermion sign problem, by building a representation of the phase fluctuations in the Boltzmann factor. This representation can be an approximate analytical expression (Section~\ref{sec:analytic}) or a neural network trained via gradient descent (Section~\ref{sec:learning}). In practice, at least with the specific methods used in this paper applied to the Thirring model, this method is outperformed by previously studied contour deformation-based methods~\cite{Alexandru:2018fqp,Alexandru:2018ddf}. The two methods can be combined, yielding a higher average phase than either alone can achieve (Section~\ref{sec:deformations}); however, this imposes a large cost for a relatively small benefit (compared to contour deformations alone).

The machine learning-based method used in this paper comprises a search for a specific object---termed a subtraction---which can be proven to exist. This contrasts sharply with the situation respecting contour deformations, where contour deformations completely solving the sign problem likely do not exist for many models (but establishing this fact rigorously is difficult even for simple models). A notable computational advantage of the method we propose compared to the contour deformation methods is that our method needs no computation of a Jacobian determinant --- the Boltzmann factor is modified to alleviate the sign problem without performing a change of variables.

It is also worth noting that the idea of modifying a Boltzmann factor by a subtraction assumes nothing about the structure of the space of field configurations. In particular, contour deformations cannot easily be applied to models with discrete field values, since there is no natural complex structure on (an extension of) field space. By contrast, finite differences can be defined on a discrete space of field configurations, and the discrete equivalent of the divergence theorem allows the methods of this paper to be applied. The task of demonstrating this, we leave to future work.

In Section~\ref{sec:deformations} we noted that the histogram of $\Im S$ on the optimal contour defined by \eq{smile} indicates that there are small deformations of that contour which improve the average phase. It may therefore be profitable to explore a deep-NN ansatz for contour deformations in the Thirring model.

Finally, in this work we showed that the average phase improves substantially as the size of the network is increased. This is in sharp contrast with many efforts to find contour deformations, where nearly all of the improvement in the sign problem was found to stem from a small number of parameters~\cite{Alexandru:2017czx,Giordano:2022miv}. However, the improvement in average phase reached a clear plateau at three or four nonlinear layers. It may be that this plateau can be lifted with more advanced methods, for example exploiting translational symmetry, using sparser networks to speed up training, or using an entirely different ansatz (such as neural ODEs~\cite{chen2018neural}).

\begin{acknowledgments}
S.L.~is grateful to Tanmoy Bhattacharya, Rajan Gupta, Jun-Sik Yoo, and Boram Yoon for many useful discussions, and for the hospitality of the Los Alamos National Laboratory. Y.Y.~appreciates the hospitality of Paul Romatschke and the University of Colorado at Boulder.

S.L.~is supported by the U.S.~Department of Energy under Contract
No.~DE-SC0017905. Y.Y.~was supported by the U.S.~Department of Energy under
Contract No.~DE-FG02-93ER-40762, and subsequently by the U.S. Department of Energy under Contract No.~DE-FG02-00ER41132.
\end{acknowledgments}

\appendix
\section{Calculation of functional derivatives}\label{app:deriv}
In this appendix we sketch the derivation of equations (\ref{sub-u}) and (\ref{sub-v}). We begin with Eq.~(\ref{sub-u}). In order to take the functional derivative it is easiest to first manipulate the quenched partition function into a slightly different form. Noting that $\nabla \cdot f u = u \cdot \nabla f + f \nabla \cdot u$, observe that Eq.~(\ref{eq:Zqu}) may be rewritten
\begin{equation}
Z_{Q,u} =
\int |e^{-S}| \left[
1 + \Re \left(u \cdot \nabla i \Im S\right) + O(u^2)
\right]
\text.
\end{equation}
We are therefore interested in the functional derivative, with respect to $u$, of $\left[\Re \int |e^{-S}| u \cdot \nabla i \Im S\right]$.

This functional derivative refers to a direction of steepest ascent, rather than a functional equivalent of the holomorphic (i.e.~Wirtinger) derivative. To avoid confusion, we may explicitly split the vector field into its real and imaginary parts $u = u_R + i u_I$ and differentiate with respect to those individually, finding that
\begin{eqnarray}
\frac{\delta}{\delta u_R} \left[\Re \int |e^{-S}| u_R \cdot \nabla i \Im S\right] &=& 0 \text{, and}\\
\frac{\delta}{\delta u_I} \left[\Re \int |e^{-S}| i u_I \cdot \nabla i \Im S\right] &=& - \nabla \Im S
\text.
\end{eqnarray}
Combining the two yields Eq.~(\ref{eq:Zqu}) as the direction of steepest ascent.

The unscaled subtraction of Eq.~(\ref{sub-v}) is derived in much the same way. Beginning with Eq.~(\ref{eq:Zqv}) and rewriting, as above, to isolate $v$ without any derivatives, we obtain
\begin{equation}
Z_{Q,v} = \int |e^{-S}| + \Re \int e^{i \Im S} v \cdot \nabla i \Im S + O(v^2)
\text.
\end{equation}
Splitting $v$ into real and imaginary parts, differentiating, and recombining as before, we obtain Eq.~(\ref{sub-v}).

\bibliographystyle{apsrev4-2}
\bibliography{References}

\end{document}